\documentclass[prl,twocolumn]{revtex4}
\usepackage{epsfig}
\usepackage{bm}
\usepackage{amsmath}

\begin{document}

\title{Helical distributed chaos in magnetic field of solar wind}

\author{A. Bershadskii}

\affiliation{
ICAR, P.O. Box 31155, Jerusalem 91000, Israel
}

\begin{abstract}

 Helical distributed chaos in magnetic field has been studied using results of direct numerical simulations (dominated by magnetic helicity), of a laboratory experiment with plasma wind tunnel and of solar wind measurements (dominated by combined magnetic and cross helicity effects). The solar wind measurements, used for the spectral computations, were produced by Helios-1 and Ulysses missions for low and high heliolatitudes respectively and for high solar wind speed at $0.4 < R <4.5$ AU (where $R$ is distance from the Sun).
\end{abstract}

\maketitle

\section{Introduction}

  Observations in solar wind, unlike numerical simulations and laboratory experiments, provide reliable scaling (power law) spectra in abundance. Alfv\'enic waves and turbulence are considered as the main sources for these spectra, although appearance of the scaling spectra in the solar wind is often unexpected and hard to explain (see, for instance, Refs. \cite{gold}-\cite{bc} and references therein). The chaotic processes, on the other hand, are characterized mainly by exponential-like power spectra \cite{fm}-\cite{mm} and, therefore, are not considered as a rule for interpretation of the solar wind observations. Taking into account the multi-scale nature of the solar wind processes it is difficult to understand, especially for large-scale phenomena. \\

   When one thinks about chaotic dynamics first thing that should be taken into account is the dynamic invariants. In ideal magnetohydrodynamics the most important for our consideration dynamic invariants are the magnetic and cross helicity: $h_m = \langle {\bf A} \cdot {\bf B} \rangle$ and  $h_{cr} = \langle {\bf v} \cdot {\bf B} \rangle$, respectively (where ${\bf B} = \nabla \times {\bf A}$ is the magnetic field strength, ${\bf A}$ is  the vector potential, ${\bf v}$ is the velocity field and $\langle ... \rangle$ denotes spatial average). In the non-ideal cases the $h_m$ and $h_{cr}$ can be still considered as adiabatic invariants for the inertial range scales, for instance.

\section{Distributed chaos and magnetic helicity}

   In the fluids and plasmas dynamics deterministic chaos at the onset of turbulence is often related to spatial exponential power spectra \cite{mm},\cite{kds}
$$
E(k) \propto \exp-(k/k_c)  \eqno{(1)}
$$       
where $k_c$ is a constant. On the way from deterministic chaos to developed turbulence the parameter $k_c$ in the Eq. (1) becomes fluctuating and one should use ensemble averaging
$$
E(k) \propto \int_0^{\infty} P(k_c) \exp -(k/k_c)dk_c, \eqno{(2)}
$$    
with a probability distribution $P(k_c)$, in order to compute the power spectra. To find $P(k_c)$ we can use a scaling relationship between characteristic value of magnetic field strength $B_c$ and $k_c$
$$
B_c \propto |h_m|^{1/2} k_c^{1/2}   \eqno{(3)}
$$
obtained employing the dimensional considerations. If the $B_c$ has Gaussian distribution (with zero mean, see below), then $k_c$ is the chi-squared ($\chi^{2}$) distributed quantity
$$
P(k_c) \propto k_c^{-1/2} \exp-(k_c/4k_{\beta})  \eqno{(4)}
$$
where $k_{\beta}$ is a constant. 

   Substituting Eq. (4) into Eq. (2) one obtains
$$
E(k) \propto \exp-(k/k_{\beta})^{1/2}  \eqno{(5)}
$$
This is magnetic energy spectrum in the inertial range of scales (see Introduction) for the magnetic helicity dominated distributed chaos. One can see that in this case the power-law spectrum in the inertial range (or at least in a part of it, see below) is replaced by a stretched exponential one.\\

  For large scales the kinetic plasma effects become non-significant and the magnetohydrodynamics can be considered as a first order approximation. In recent paper Ref. \cite{bl} results of direct numerical simulations (DNS) of decaying magnetohydrodynamic turbulence were reported. 
  
   The magnetohydrodynamic equations for an incompressible fluid (in the Alfv\'enic units) 
$$
 \frac{\partial {\bf u}}{\partial t} = - ({\bf u} \cdot \nabla) {\bf u} 
    -\frac{1}{\rho} \nabla {\cal P} - [\tilde{{\bf B}} \times (\nabla \times \tilde{{\bf B}})] + \nu \nabla^2  {\bf u} \eqno{(6)}
$$
$$
\frac{\partial \tilde{{\bf B}}}{\partial t} = \nabla \times ( {\bf u} \times
    \tilde{{\bf B}}) +\eta \nabla^2 \tilde{{\bf B}}    \eqno{(7)} 
$$
were solved (numerically) in a cubic domain with usually used periodic boundary conditions. In the Alfv\'enic units the magnetic field strength $\tilde{{\bf B}}={\bf B}/\sqrt{\mu_0\rho}$ has the same dimension as the velocity field. 

  In this DNS a background magnetic field was not imposed and the initial velocity and magnetic fields were random Gaussian (with zero mean). The initial relative magnetic helicity was normalized by magnetic energy to value 1, while the initial cross helicity and the velocity based helicity were negligible. The initial Taylor-Reynolds number $R_{\lambda} (0) = 74.84$.
  
  Figure 1 shows magnetic energy spectrum for the time of the decay equal to one initial turnover time (the spectral data were taken from Fig. 1 of the Ref. \cite{bl}). The dashed curve indicates correspondence to the stretched exponential spectrum Eq. (5).

\section{Distributed chaos and cross helicity}

  While the magnetic helicity is an indicator of the lack of reflectional symmetry and is usually related to the dynamo effects, the cross helicity can be considered as a measure of relative importance of the Alfv\'en waves.
  
   One cannot use the cross helicity to obtain relationship between $B_c$  and $k_c$ as it was done above with the magnetic helicity (cf. Eq. (3)) even if one will use the Alfv\'enic units $\tilde{{\bf B}}={\bf B}/\sqrt{\mu_0\rho}$ (with $\tilde{{\bf B}}$ having the same dimension as ${\bf v}$). Therefore, let us use a combined dynamic invariant
$$
I = \tilde{h}_{cr} \tilde{h}_m  \eqno{(7)}
$$
where $\tilde{h}_{cr} = \langle {\bf v} \cdot \tilde{{\bf B}} \rangle$ and $\tilde{h}_m =\langle \tilde{{\bf A}} \cdot \tilde{{\bf B}} \rangle $ are the cross and magnetic helicity in the Alfv\'enic units. Then we can write
$$
\tilde{ B}_c \propto |I|^{1/4} k_c^{1/4}  \eqno{(8)}
$$
instead of the Eq. (3) in this case. 
\begin{figure} \vspace{-2cm}\centering
\epsfig{width=.45\textwidth,file=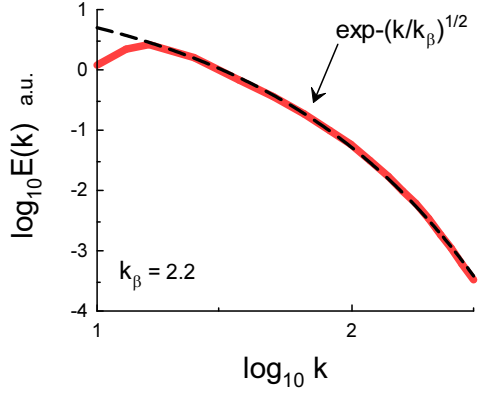} \vspace{-4cm}
\caption{ Magnetic energy spectrum for the time of the decay equal to one initial turnover time. } 
\end{figure}
\begin{figure} \vspace{+5.2cm}\centering
\epsfig{width=.8\textwidth,file=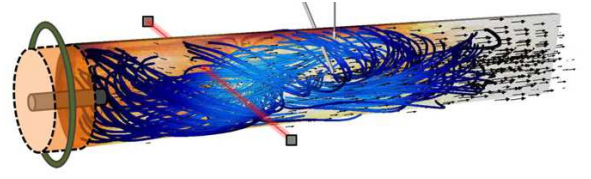} \vspace{-10cm}
\caption{A diagram  of the MHD wind tunnel.} 
\end{figure}
\begin{figure} \vspace{-0.45cm}\centering
\epsfig{width=.45\textwidth,file=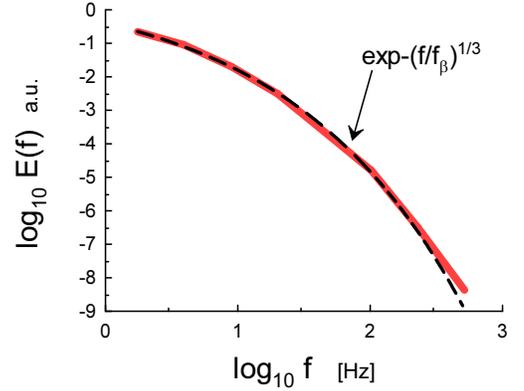} \vspace{-4.15cm}
\caption{Total magnetic energy frequency spectrum measured in the experiment.} 
\end{figure}
   If we will try the stretched exponential spectrum in this case as well 
$$
E(k) \propto \int_0^{\infty} P(k_c) \exp -(k/k_c)dk_c  \propto \exp-(k/k_{\beta})^{\beta},  \eqno{(9)}
$$
then from the Eq. (9) the asymptote of the $P(k_c)$ at large $k_c$ can be estimated as \cite{jon}
$$
P(k_c) \propto k_c^{-1 + \beta/[2(1-\beta)]}~\exp(-bk_c^{\beta/(1-\beta)}) \eqno{(10)}
$$       
 
   For the Gaussian $\tilde{ B}_c$ one obtains from the Eqs. (8) and (10) value of $\beta =1/3$, i.e.:
$$
E(k) \propto \exp-(k/k_{\beta})^{1/3}  \eqno{(11)}
$$
  
   In recent paper Ref. \cite{sbl} results of a laboratory magnetic turbulent plasma experiment in MHD wind tunnel were reported. The experiment was designed to model the solar wind processes. A set amount of magnetic helicity was generated and governed by the initial conditions. 

   Figure 2 (adapted from the Ref. \cite{sbl}) shows a diagram  of the MHD wind tunnel with a plasma gun. Orange color intensity indicates electron density, whereas the blue lines indicate simulated magnetic field twisting
under the magnetic helicity conservation. Figure 3 shows total magnetic energy frequency spectrum (ensemble averaged) measured in the experiment (the spectral data were taken from Fig. 15a of the Ref. \cite{sbl}). The Taylor hypothesis relates the frequency spectrum $E(f)$ to analogous wavenumber one-dimensional spectrum $E(k)$ by transformation $f = V_0k/(2\pi)$, where $V_0$ is the constant mean velocity of the plasma along the wind tunnel (see, for instance, recent Ref. \cite{tbn} and references therein). The dashed curve indicates correspondence to the stretched exponential spectrum Eq. (11).
   
\section{Distributed chaos in solar wind }

   The Helios-1 and Ulysses measurements were made at low and high heliolatitudes respectively. Therefore, despite of the high variability of the solar wind properties, a comparison of the results obtained from these measurements can provide a general picture (see, for instance, Ref. \cite{hb} and references therein). A preliminary picture obtained from the measurements of magnetic field fluctuations in the solar wind consists of the large scale structures, Alfv\'en waves at intermediate scales and small scales turbulent fluctuations between 0.3 (Helios-1) and 4.1 (Ulysses) astronomical units (AU) from the Sun, showing a strong similarity for the Helios-1 and Ulysses data in the high-speed streams \cite{hb}. \\
\begin{figure} \vspace{-3cm}\centering
\epsfig{width=.45\textwidth,file=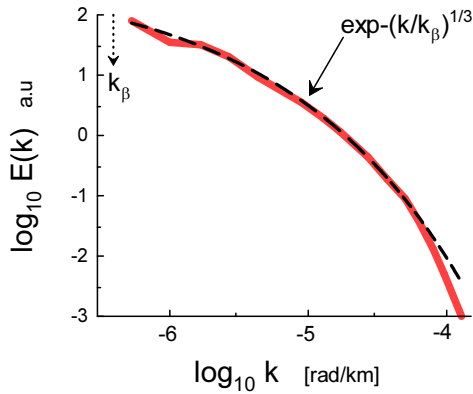} \vspace{-2.5cm}
\caption{Total magnetic energy spectrum computed using measurements made by Helios-1.} 
\end{figure}

\begin{figure} \vspace{-1.3cm}\centering
\epsfig{width=.45\textwidth,file=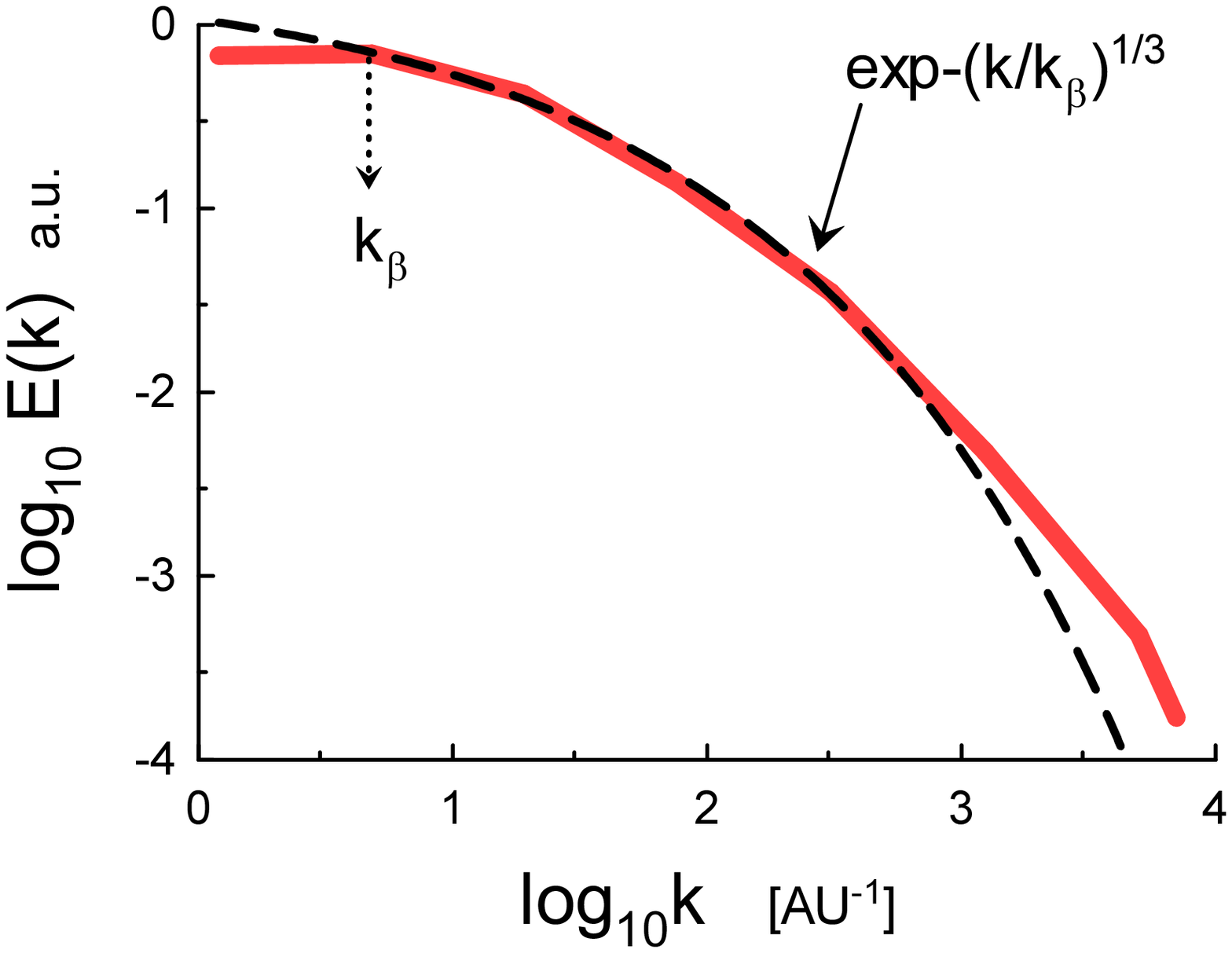} \vspace{-4.2cm}
\caption{Total magnetic energy spectrum computed using measurements made by Ulysses for $1.5 < R < 2.8$ AU.} 
\end{figure}

\begin{figure} \vspace{-0.5cm}\centering
\epsfig{width=.45\textwidth,file=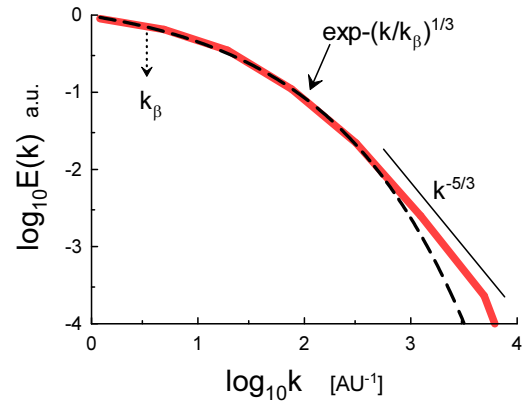} \vspace{-4.2cm}
\caption{Total magnetic energy spectrum computed using measurements made by Ulysses for $2.8 < R < 4.5$ AU.} 
\end{figure}
   Figure 4 shows total magnetic energy spectrum computed using measurements made by Helios-1 from March 3,
1975, 1200UT to March 4, 1975, 1200UT (the spectral data were taken from Fig. 3 of the Ref. \cite{tbn} - full speed mapping). The Helios-1 spacecraft was located at distance $R=0.4$ AU from the Sun. \\

  The dashed curve in the Fig. 4 indicates correspondence to the stretched exponential spectrum Eq. (11). The dotted arrow indicates position of the scale $k_{\beta}$. One can see that in this case (as in the case of the MHD wind tunnel experiment, cf. Fig. 3) the combined magnetic and cross helicity domination, Eqs. (7-8), takes place. Let us recall that the cross helicity can be considered as a measure of relative importance of the Alfv\'en waves (see, for instance, Ref. \cite{rob} for a possible role of the cross helicity in the solar wind dynamics for the distances 0.3 to 5 AU from the Sun).\\
  
  Figure 5 shows total magnetic energy spectrum computed using measurements made by Ulysses for 1993-1996yy period at high heliolatitudes and at high solar wind speed (the spectral data were taken from Fig. 3 of the Ref. \cite{bran} for $1.5 < R < 2.8$ AU, where $R$ is distance from the Sun). The spectrum was rescaled by factor $4\pi R^2$ before averaging over the data sets. Figure 6 shows analogous spectrum but for $2.8 < R < 4.5$ AU (the spectral data were taken from Fig. 3 of the Ref. \cite{bran}). The dashed curves indicate correspondence to the stretched exponential spectrum Eq. (11) (cf. Figs. 3,4).\\
  
  Finally let us discuss briefly situation at kinetic (electron) scales. At $R \sim 1$ AU the electron spatial scales are
of order of a few kilometres (in the frequency domain between 30 and 300 $Hz$). Authors of the Ref. \cite{Cluster}, analysing the corresponding data obtained by the Cluster mission \cite{esg}, suggested an empirical fit for the magnetic power spectrum in the stretched exponential form: $E(f) \propto \exp-(f/f_{\beta})^{1/2}$ (cf. Eq. (5)). However, it is not clear whether the methodology developed above for the MHD scales can be applied to the kinetic scales investigated by the Cluster mission as well \cite{gold2}.

\end{document}